\documentclass[%
reprint,
superscriptaddress,
 amsmath,amssymb,
aip,
]{revtex4-2}

\usepackage{graphicx}% Include figure files
\usepackage{dcolumn}% Align table columns on decimal point
\usepackage{bm}% bold math
\usepackage{pifont}
\usepackage[version=3]{mhchem} % Formula subscripts using \ce{}
\usepackage{braket}
\usepackage{float}
\usepackage{mathtools}

\DeclareMathSymbol{*}{\mathbin}{symbols}{"03} % \ast
\DeclareMathSymbol{\ast}{\mathbin}{symbols}{"03}

\usepackage{tikz}

\begin{document}

\preprint{APS/123-QED}
\title{A parameter-free statistical model  for two-dimensional carbon nanostructures}% Force line breaks with \\

\author{Chang-Chun He}
\thanks{These authors contributed equally.}
\affiliation{School of Physics and Optoelectronics, South China University of Technology, Guangzhou 510640, China}
\author{Shao-Gang Xu}
\thanks{These authors contributed equally.}
\affiliation{Department of Physics, Southern University of Science and Technology, Shenzhen 518055, China}
\affiliation{Quantum Science Center of Guangdong-Hong Kong-Macao Greater Bay Area (Guangdong), Shenzhen 518045, People’s Republic of China}
\author{Jiarui Zeng}
\affiliation{School of Physics and Optoelectronics, South China University of Technology, Guangzhou 510640, China}
\author{Weijie Huang}
\affiliation{School of Physics and Optoelectronics, South China University of Technology, Guangzhou 510640, China}
\author{Yao Yao}
\affiliation{School of Physics and Optoelectronics, South China University of Technology, Guangzhou 510640, China}
\affiliation{State Key Laboratory of Luminescent Materials and Devices, South China University of Technology, Guangzhou 510640, China}
\author{Yu-Jun Zhao}
\affiliation{School of Physics and Optoelectronics, South China University of Technology, Guangzhou 510640, China}
\author{Hu Xu}
\email{xuh@sustech.edu.cn}
\affiliation{Department of Physics, Southern University of Science and Technology, Shenzhen 518055, China}
\affiliation{Quantum Science Center of Guangdong-Hong Kong-Macao Greater Bay Area (Guangdong), Shenzhen 518045, People’s Republic of China}
\author{Xiao-Bao Yang}
\email{scxbyangscut@scut.edu.cn}
\affiliation{School of Physics and Optoelectronics, South China University of Technology, Guangzhou 510640, China}

\date{\today}% It is always \today, today,

\begin{abstract}
Energy degeneracy in physical systems may be induced by symmetries of the Hamiltonian, and the resonance of  degeneracy states in carbon nanostructures can effectively enhance the stability of the system. Combining the octet rule,  we introduce a parameter-free statistical model to determine the physical properties by lifting the energy degeneracy in carbon nanostructures. This model offers a direct path to accurately ascertain electron density distributions in quantum systems, akin to how charge density is used in density functional theory  to deduce system properties. Our methodology diverges from traditional quantum mechanics, focusing instead on this unique statistical model by minimizing bonding free energy to determine the fundamental properties of materials. Applied to carbon nanoclusters and graphynes, our model not only precisely predicts bonding energies and electron density without relying on external parameters,  but also enhances the prediction of electronic structures through bond occupancy numbers, which act as effective hopping integrals. This innovation offers insights into the structural properties and quantum behavior of electrons across various dimensions.
\end{abstract}

\maketitle

\section{Introduction}
Energy degeneracy resulting from latent  symmetry is a crucial aspect of quantum systems, contributing to various  quantum phenomena  \cite{RevModPhys.35.1,vonKlitzing2020,10.1063/1.1654509}. The latent symmetry is attributed to the symmetry of the isospectral effective Hamiltonian obtained through subsystem partitioning, and the rotational symmetries can be broken in a controlled manner while maintaining the more fundamental symmetry   \cite{PhysRevLett.126.180601}.  Lifting the degeneracy by external perturbations can lead to a lower energy state with energy level  splitting.  For instance, the Jahn-Teller effect describes how local geometric distortions can lower the overall energy of a system by removing degeneracy   \cite{doi:10.1098/rspa.1937.0142}. In carbon nanostructures, quantum resonance    provides another mechanism for lifting energy degeneracy by the combination of possible degenerate Kekul$\acute{\textrm{e}}$ structures,  emerging delocalized electron states and aromaticity  to further stabilize   the system    \cite{Muller+1994+1077+1184,C5CS00183H,Wang2017_nat_chem} For the benzene molecule,  there are two  Kekul$\acute{\textrm{e}}$ structures  with C$_3$ symmetry that both meet the octet rule, exhbiting the two-fold energy degeneracy. Resonance theory shows that  the uniform superposition of  Kekul$\acute{\textrm{e}}$ structures  with C$_6$ symmetry  corresponds to the true electron density of benzene   \cite{Liu2020_nat_comm},   lifting the energy degeneracy with additional resonance energy.

%For graphene nanoribbons,  different atomic spacing along the armchair and zigzag edges results in distinctly different electron density distribution   \cite{PhysRevLett.105.235502}.
% 加上hopping系数的拟合，由于应变导致hopping的变化
% clar 结构提供了一种建议的去简并的方式
%PAHs are a class of graphene nanoflakes characterized by strong in-plane $\pi$-$\pi$ conjugation   \cite{C5CS00183H,Wang2017_nat_chem}, and possess a network of delocalized $\pi$ electrons. 

Generally, a  polycyclic aromatic hydrocarbon (PAH) molecule has a series of Kekul$\acute{\textrm{e}}$ structures that satisfy the octet rule in degenerate energy  state.  To construct the most valuable one to represent all  Kekul$\acute{\textrm{e}}$ structures,  Clar's rule identifies the key structure by maximizing the number of nonadjacent $\pi$-sextets   \cite{Clar1983}.  The nonadjacent  $\pi$-sextets  have two  forms with alternating single and double bonds,  independently contributing a statistical weight of 2, and  the statistical weight is $2^N$ for $N$ $\pi$-sextets. Therefore,  Clar's rule provides an avenue to lift energy degeneracy  by favoring the structure with the largest statistical weight,  which best approximates the relative stability of various carbon nanostructures   \cite{PhysRevLett.97.216803,PhysRevLett.119.076401,doi:10.1021/cr9903656, doi:10.1021/ja909234y}.   The sextets predicted by Clar's rule generally correspond to aromatic rings, aligning with various theoretical methods   \cite{doi:10.1021/ci050196s, doi:10.1021/jp040179q}, though exceptions exist in irregular-shaped PAHs   \cite{doi:10.1021/acs.jpca.6b00972}. Additionally, PAH isomers with more sextets tend to exhibit higher kinetic stability, lower reactivity, and larger energy gaps   \cite{https://doi.org/10.1002/chem.201905087}. Despite  the success in interpreting chemical bonding,  the empirical rules limit  quantitative electron distribution analysis for lack of rigorous mathematical foundation in quantum theory   \cite{Wang2022_jpca,doi:10.1021/acs.jcim.1c00735,gutman2013test}. A single Clar structure often cannot accurately reproduce the actual electron density distribution of a PAH molecule, as the symmetry of the Clar structure often disagrees with the realistic molecular symmetry   \cite{CYVIN1986859}. % The resonance theory describes chemical bonding by combining several key contributing structures with specific weights into a resonance hybrid   \cite{doi:10.1021/ja00828a015}, where the weights are derived  using $ab~initio$ methods derived from valence bond theory   \cite{doi:10.1021/j100017a027}  or else from the natural bond orbitals  approaches   \cite{Glendenin_1998}.  % To best approximate the actual wave function, a superposition of all possible Clar resonators is constructed, where the weights of the resonators are determined through a projection-weighted symmetric orthogonalization scheme  from the density functional theory (DFT) wave function   \cite{Wang2022_jpca}. 

The Clar's rule, as a parameter-free  model,  falls short in accurately predicting the electron density along with structural stabilities and electronic properties of carbon nanostructures. Current quantitative models require training data from DFT to acquire model parameters and precisely predict the properties of unknown structures, which belong to a first-principles-based (second-principles) scheme   \cite{PhysRevB.93.195137}.   To address electronic properties, the tight-binding (TB) model offers a simpler method to construct an effective Hamiltonian   \cite{PhysRevB.39.12520}, but it relies on DFT or experimental data for integral parameters   \cite{C7RA06891C}.  Several deep learning methods have been proposed to predict structural properties after fitting millions of  parameters based on massive data from  DFT   \cite{Chandrasekaran2019,Jrgensen2022_nat_comm,delRio2023}. However, the inherent chemical origin and deeper insights into specific systems are obscured by the intricate neural network model. Therefore, there is a need for a more direct parameter-free method that can both conceptualize and quantify the electron density distribution, providing a deeper understanding of chemical bonding in carbon nanostructures.

In this work,  we introduce a streamlined parameter-free statistical  bonding free energy (BFE) model   for determining electron density distribution  by lifting the energy level in carbon nanostructures. The predicted structural properties of this model remarkably align with both experimental findings and first-principles calculations.  By integrating the octet rule with the BFE model within a grand canonical  ensemble, we can accurately  evaluate bonding energies, uncovering significant contributions of bonding entropy to structural stability in carbon nanostructures. Furthermore, the  occupancy numbers (ONs) of each C-C bond derived from our model predict   the electron density and bond length.  These ONs can serve as hopping integrals in the TB models, thereby simplifying accurate predictions of electronic structures, including energy gaps,  energy levels, and the corresponding   spatial distribution of wavefunctions. Local aromaticity  is defined by the  mean value of ONs in   each six-membered ring, offering a simple yet effective method to measure  electron delocalization in the $\pi$-conjugated  system. Our method provides valuable insights into the structural properties and quantum behavior of electrons across different dimensional carbon nanostructures.
%\newpage
\section{RESULTS AND DISCUSSION}
\subsection{The derivation of bonding free energy model}
\begin{figure}[ht]
    \centering
    \includegraphics[width=\linewidth]{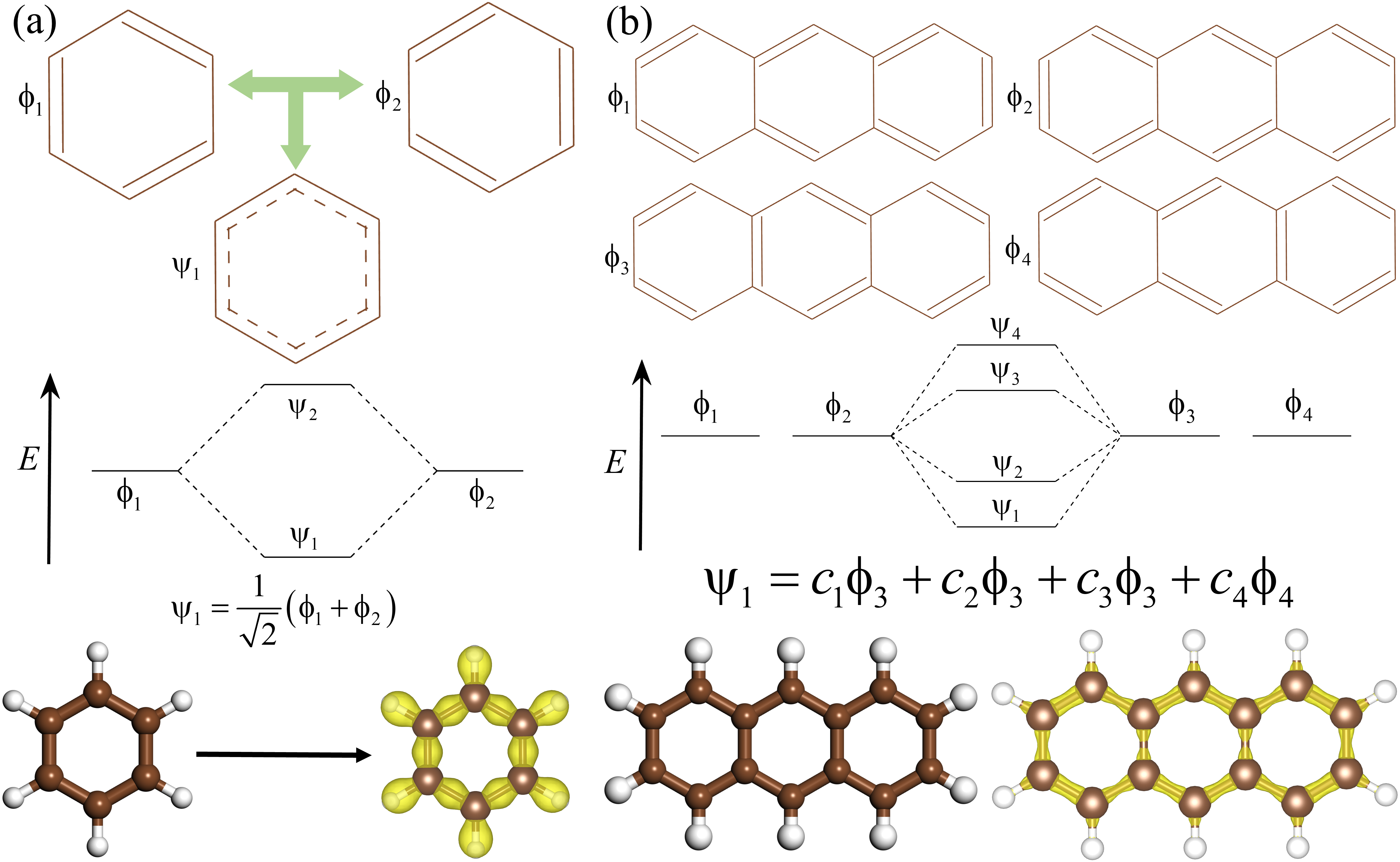}
    \caption{(a) Top part is two Kekul$\acute{\textrm{e}}$ structures (denoted by $\phi_1, \phi_2$) and the average structure of benzene (denoted by $\psi_1$); middle part is  the energy level diagram of two degenerate state $\phi_1, \phi_2$; bottom part is the structure and charge distribution of benzene calculated by DFT. (b) Top part is  four Kekul$\acute{\textrm{e}}$ structures of anthracene (denoted by $\phi_1$-$\phi_4$); middle part is the   energy level diagram of four degenerate state $\phi_1$-$\phi_4$ with the undetermined parameters;  bottom part is the structure and charge density of anthracene calculated by DFT. }
    \label{fig1}
\end{figure}

To understand the interatomic interactions in carbon nanostructures, we start from the simple benzene molecule, which has two different forms of alternating single and double bonds that obey the octet rule as shown  in the top panel of FIG. 1(a). The degenerated Kekul$\acute{\textrm{e}}$ structures hold three-fold symmetry, where the fictional C-C double bond is stronger than C-C single bond, violating the  identical C-C bonds in the benzene molecule.  To lift the two-fold degenerate energy level,  the equally weighted superposition of two Kekul$\acute{\textrm{e}}$ structures can satisfy the intrinsic symmetry as depicted in the middle panel of FIG. \ref{fig1}(a), which has the lowest energy due to the additional resonance energy.     From the view of molecular theory, the optimal state $\psi_1$ is the resonance of  two equal Kekul$\acute{\textrm{e}}$ structures,  thus  the resonance weight is $1/\sqrt{2}$ to ensure that the ONs of all C-C bonds are identical to 1.5, in agreement with the  charge density distribution derived from DFT  as shown in bottom part of  FIG. \ref{fig1}(a).  For any carbon nanostructures, the ground state  can be represented by the linear combinations of a series of degenerate Kekul$\acute{\textrm{e}}$ structures  \cite{https://doi.org/10.1002adsasf}, but the resonance weights can not  directly derived unless from   first-principles calculations. For instance, the anthracene has four Kekul$\acute{\textrm{e}}$ structures (see FIG. \ref{fig1}(b)). Particularly,   the charge density from DFT  shows that the edge C-C bonds are stronger than the inner C-C bonds  in the bottom part of FIG. \ref{fig1}(b), but we can not determine the accurate  ONs of C-C bond, while the proposed BFE model can easily calculate the resonance weights with external parameters.

\begin{figure}
    \centering
    \includegraphics[width=\linewidth]{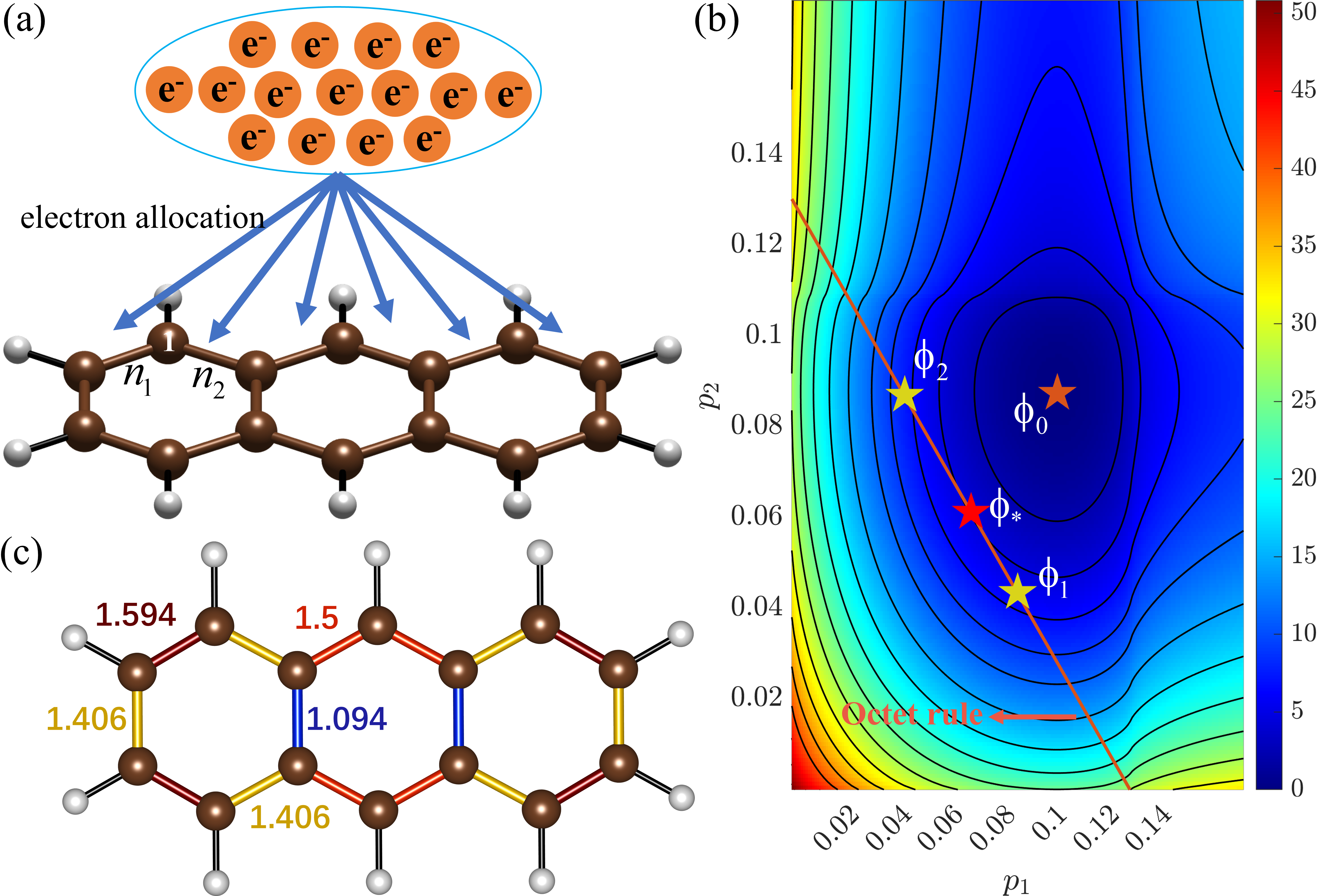}
    \caption{(a) Total electron allocation model for anthracene molecule. (b) The bonding free energy surface varied with $p_1$ and $p_2$ for anthracene molecule. (c) The optimal occupancy number of each bond for anthracene molecule, and the ON number is marked near the bond.}
    \label{fig2}
\end{figure}

For a given  carbon nanostructure, a series of possible degenerate Kekul$\acute{\textrm{e}}$ structures satisfy the octet rule. Assuming that all electrons are mobile  across all chemical bonds and the number of electrons in each bond is variable, the system can be considered as  a grand canonical ensemble.   In this framework, the total electrons in the system are allocated across each bond as demonstrated in FIG. \ref{fig2}(a),   where  each bond contains $n_i$ electrons, associated with a chemical potential $\mu_i$, and upholds the total electron constraint $\sum_{i=1}^{N_{\textrm{bond}}} n_i=N_{\textrm{ele}}$. Here, $N_{\textrm{bond}}$ and $N_{\textrm{ele}}$ denote the total numbers of bonds and electrons, respectively. Assuming that the internal energy for C-C bonds is equivalent,   the  chemical potential $\mu_i$ can be used to distinguish nonequivalent C-C bonds, with internal energy $U$ set to zero for simplicity. The grand canonical partition function is derived by
\begin{equation}
  {\cal{Z}} =  \smashoperator{\sum_{n_1,...,n_{N_{\textrm{bond}}}}} e^{-\sum_{i=1}^{N_{\textrm{bond}}} (n_i\alpha_i +U)} = \left(\sum_{i=1}^{N_{\textrm{bond}}}e^{-\alpha_i}\right)^{N_{\textrm{ele}}},
\end{equation}
where $\alpha_i = -\frac{\mu_i}{k_{\textrm{B}} T}$ is the reduced chemical potential. The first equality encompasses the ergodic combinations of electron numbers for each bond   \cite{PhysRevLett.87.250601,PhysRevLett.122.010601},  including all  possible Kekul$\acute{\textrm{e}}$ structures. The second equality holds because all combinations correspond to a multinomial expansion, where the multinomial coefficients correspond to the multiplicity of each  combination. We then express the formula for free energy as:
\begin{equation}
    F = N_{\textrm{ele}}k_{\textrm{B}}T   \sum_{i=1}^{N_{\textrm{bond}}} p_i \log p_i = -N_{\textrm{ele}}k_{\textrm{B}}T S_{\textrm{b}},
\end{equation}
where the  bonding entropy $S_{\textrm{b}}$ is $-\sum_{i=1}^{N_{\textrm{bond}}} p_i \log p_i$. $p_i$ is the electron probability in bond $i$, defined by 

\begin{equation}
   p_i = \frac{n_i}{N_{\textrm{ele}}} = -\frac{1}{N_{\textrm{ele}}}\frac{\partial F}{\partial \mu_i} = \frac{e^{-\alpha_i}}{\sum_{i=1}^{N_{\textrm{bond}}} e^{-\alpha_i}}.
\end{equation}

It is significant to note that the  energy is roughly  proportional to the total number  of electrons $N_{\textrm{ele}}$ in the system, and the bonding entropy is proportional to $\log N_{\textrm{ele}}$. We introduce the concept of an equivalent ``temperature'' to redefine the bonding free energy:
 
\begin{equation}
\begin{split}
    F_{\textrm{b}}(p_1,...,p_{N_\textrm{bond}})  & =   \frac{N_{\textrm{ele}}k_{\textrm{B}}T_0}{\log N_{\textrm{ele}}} \sum_{i=1}^{N_{\textrm{bond}}}p_i\log(p_i),
\end{split}
    \label{free_BE}
\end{equation}
where the equivalent ``temperature'' $k_{\textrm{B}}T$ is $\frac{k_{\textrm{B}}T_0}{\log N_{\textrm{ele}}}$,   $T_0$ is the equivalent ``standard temperature'', and $k_{\textrm{B}}T$ functions as the coefficient to ensure that  $F_{\textrm{b}}$ is an extensive quantity as detailed in the Supplementary Material (SM).  The $F_{\textrm{b}}$  as a function of $p_i$ is determined by Eq. \ref{free_BE}, where the minimal $F_{\textrm{b}}$ corresponding to the maximum bonding entropy $S_{\textrm{b}}$  can  decide the  ground state electron density  for carbon  nanostructures.  Here, we take  anthracene  as an example to show how to determine the $p_i$ of each bond. The variation of $F_{\textrm{b}}$  with  $\phi_1$ and $\phi_2$ is given in FIG.  \ref{fig2}(b), where the global minimum $\phi_0$ without any  constraints is marked by  the orange pentagon.  $\phi_0$ represents the equal electron probability for each bond,  which contradicts the actual electron distribution of the molecule due to varying bonding environments for each carbon atom.  

The identification of the minimal $F_{\textrm{b}}$ is essential, particularly under local bonding constraints ensuring each carbon atom adheres to the octet rule   \cite{doi:10.1021/ja01447a011,doi:10.1021/jacs.3c10370}. For instance, we apply the constraint $n_1+n_2=6$ to carbon atom 1, taking into account the two-electron requirement of the C-H bond to satisfy the duplet rule for hydrogen. FIG.  \ref{fig1}(c) shows four Kekul$\acute{\textrm{e}}$ structures  that satisfy the octet rule, with their respective $F_{\textrm{b}}$ values depicted in FIG.  \ref{fig2}(b). The pink line represents the area compliant  with the octet rule in FIG.  \ref{fig1}(b), and the red pentagon indicates the  optimal  electron density of $\phi_*$   as depicted in FIG.  \ref{fig2}(c).  The $F_{\textrm{b}}$ for $\phi_*$ distribution is notably lower than that of the Kekul$\acute{\textrm{e}}$ structures, indicating greater stability in the resonance structure compared to a single Lewis structure. Similar to bond order, the ONs of each bond, defined by $n_i/2$ for C-C bonds in two-center two-electron bonds, are indicated near each inequivalent bond. Note that the predicted ONs are in  good agreement with  the molecular symmetry  compared with the Clar structure   \cite{10.3389/fchem.2013.00022}. % Additionally, the ONs also agree with the charge density and bond length   \cite{doi:10.1098/rspa.1951.0119} as detailed in the  SM, while the Clar structure can not distinguish the bond order difference between bond 1 and bond 2 as shown in FIG. \ref{fig1}(d).  Note the ONs of phenanthrene  are more uniform  compared with that of anthracene, and  the ON of  middle C-C bond in anthracene is closer to  C-C single bond,  contributing the instability.

\subsection{Energy prediction ability of BFE model }

\begin{figure*}
    \centering
    \includegraphics[width=\linewidth]{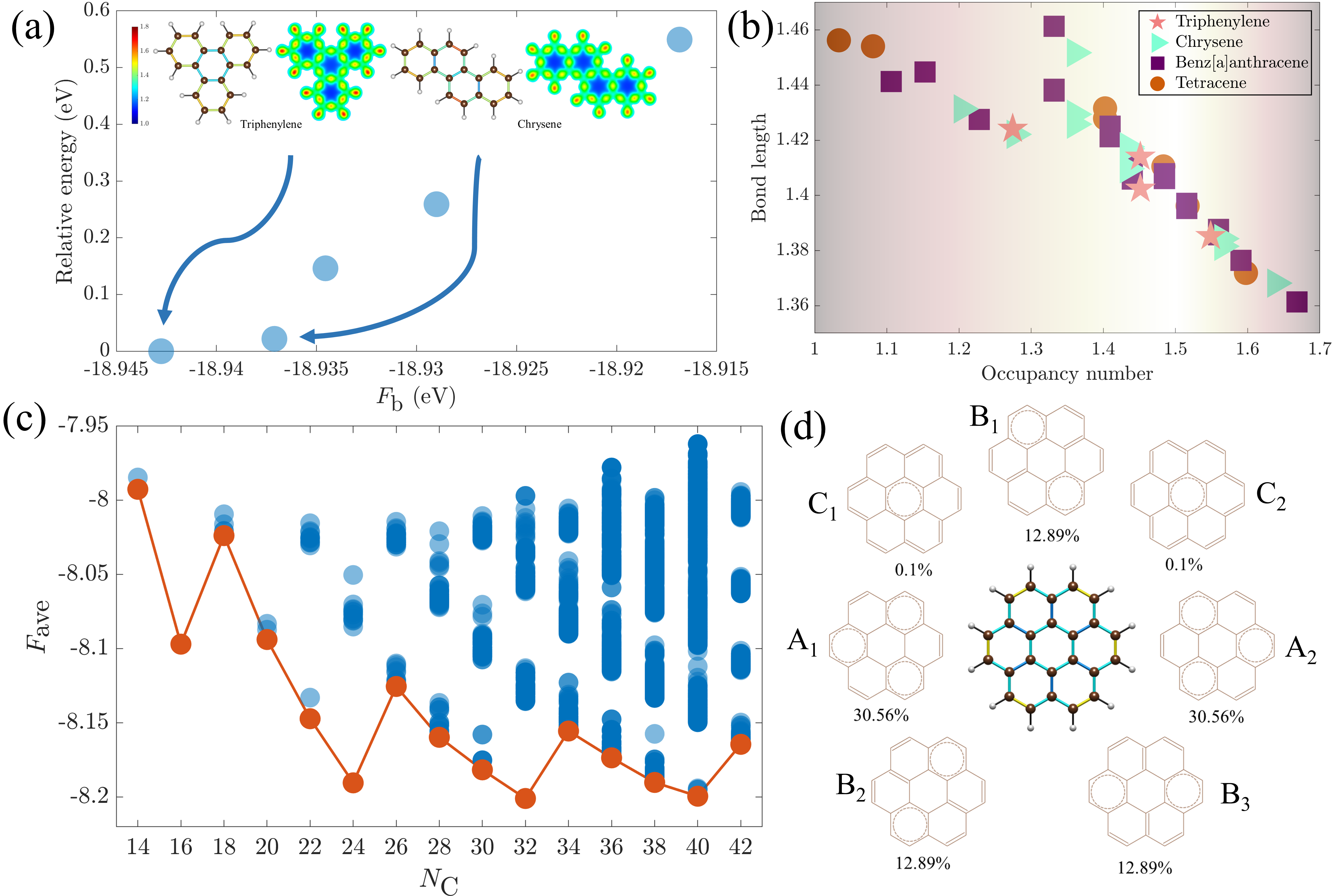}
    \caption{(a) The linear relationship between the relative energy  predicted by the DFT and $F_{\textrm{b}}$ calculated by BFE model, where the charge density and ON of two stable  \ce{C18H12} molecules are plotted in the inset.  (b) The  relationship between the ON predicted by BFE model and  bond length determined by the DFT. (c) The convex-hull of $F_{\textrm{ave}}$ predicted by BFE model, and the convex points are marked by pink circles. (d) The seven resonance structures for coronene molecule, and the derived resonance weights is marked near the structures.}
    \label{fig3}
\end{figure*}

According to the BFE model, the bond energies of carbon nanostructures can be expressed by $F_{\textrm{b}}$. Thus, the relative stability of isomers can be distinguished by global minimum  $F_{\textrm{b}}$. As reflected in FIG.  \ref{fig3}(a),  the first-principles calculations based on highly precise  CCSD(T)/CBS method  \cite{https://doi.org/10.1002/jcc.24669}   proves that  triphenylene with three-fold symmetry is the most stable among  \ce{C18H12} molecules.  Interestingly, the $F_{\textrm{b}}$ of triphenylene is lower than that of chrysene, aligning well with the CCSD(T)/CBS method. 
Further, the $F_{\textrm{b}}$ has well agreement with bonding energy calculated by DFT for the PAH structures comprising fewer than 30 carbon atoms, as shown in FIG. S1 of SM. Thus, BFE effectively describes PAH stability without external parameters, offering a distinct perspective on isomerization energies. As shown in the inset of  FIG. \ref{fig3}(a), the ONs in these structures reflect the symmetry of the molecules and align with the charge density distribution obtained from DFT, implying that the ON can reflect the local electron distribution.  Moreover, FIG.  \ref{fig3}(b) reveals a strong  linear relationship between the ON of bonds and bond length, where shorter bond lengths corresponding to higher ON values with stronger interactions. Triphenylene, which identified as the most stable molecule, has the lowest ON deviation from 1.5, whereas other molecules show significant deviations, contributing to their relative instability.

 Having established the BFE model of PAH structures, we are able to predict stable magic PAH clusters using the average BFE, $F_{\textrm{ave}} = \frac{F_{\textrm{b}}}{N_{\textrm{C}}}$, where $N_{\textrm{C}}$ is the total number of carbon atoms. As shown in FIG. \ref{fig3}(c), we generated all PAH structures comprising fewer than 42 carbon atoms. This allows us to investigate the variation of average BFE as a function of $N_{\textrm{C}}$ and the corresponding magic structures. Notably, the trend of $F_{\textrm{ave}}$ is the same as the formation energy derived from DFT calculations   \cite{C7RA06891C}, where the identified magic clusters, including   \ce{C16H10}, \ce{C24H12}, \ce{C32H14}, and \ce{C40H16} are depicted in FIG. S2 of SM. Therefore, $F_{\textrm{ave}}$ serves as a reliable metric for evaluating the structural stability and predicting the ground state structures, including the electron density distribution, which is equivalent to the DFT-based Hamiltonian.

% As shown in FIG. \ref{fig2}(d), the ON of each bond in these magic PAH clusters is clearly displayed, with a colorbar for reference. It is apparent that the bonds at the edges possess higher electron density than those in the interior. This is because edge carbon atoms, with a single C-H bond, necessitate a greater allocation of electrons to C-C bonds to fulfill the octet rule. Consequently, the ON of edge bonds approaches $1.5$, akin to the benzene, while the ON for internal bonds converges to $\frac{4}{3}$, resembling the graphene. Based on the ON of each bond, we have identified a deviation between ON and $1.5$ to distinguish separate hexagonal rings with minimal discrepancy from the benzene, referred to as $\pi$-sextets, marked by yellow circles in these structures. An arrow indicates that the $\pi$-sextet can migrate to the next position due to resonance, as supported by previous studies   \cite{10.3389/fchem.2013.00022}. %Interestingly, the number of carbon atoms in these magic PAH structures is consistently a multiple of eight, as every additional set of eight carbon atoms incorporates three more carbon rings and an extra $\pi$-sextet, leading to a symmetrical and stable PAH. %These structures can be viewed as hexagonal structures growing outward from a central benzene ring,  indicating that the BFE not only stands out as a highly suitable descriptor in elucidating structural stability  but also  providing a detailed and accurate understanding of the  initial nucleation process of graphene   \cite{PhysRevMaterials.7.084202}.

Beyond the stability evaluation, the resonance weights {$W_R$} (satisfying $W_R>0$, $\sum_R W_R=1$) of all  resonance structures can be defined by the optimal resonance-type representation of $\phi_*$ determined by the minimal BFE model \cite{doi:10.1021/jacs.8b12336}:
\begin{equation}
    W_R =  \textrm{argmin} ||\phi_{\ast}-\sum_R W_R\phi_R||,
\end{equation}
where $\phi_R$ represents the ON of resonance structures (the C-C single bond is 1, the C-C double bond is 2, and C-C partial double bond is 1.5 like benzene), and the $\phi_{\ast}$ represents the ON of the structures derived by BFE model.  The weights for  anthracene  in FIG. \ref{fig1}(b) are $c_1=c_4=40.6\%, c_2=c_3=9.4\%$. In addition, we take the coronene,  as a magic cluster in FIG. \ref{fig3}(c),  to show how to calculate the resonance weight, where the optimal electron distribution is shown in the center of FIG. \ref{fig3}(d). The seven surrounding structures are classified into three categories. The A$_1$, A$_2$, structures with three Clar rings in the left-center and right-center, as the most significant contributors, occupy the highest proportion of  61.12\%. Next are three structures (B$_1$, B$_2$, B$_3$) with two Clar rings, which  occupy about 38.67\% contribution. Lastly, the two structures (C$_1$, C$_2$) at the left-top and the right-top, each with only one Clar ring, contribute only 0.2\%. The result also  has good agreement with the previous work   \cite{Wang2022_jpca} and  Clar's rule   \cite{10.3389/fchem.2013.00022}, where the more  aromatic $\pi$-sextets will contribute higher stability.

\begin{figure}[H]
    \centering
    \includegraphics[width=\linewidth]{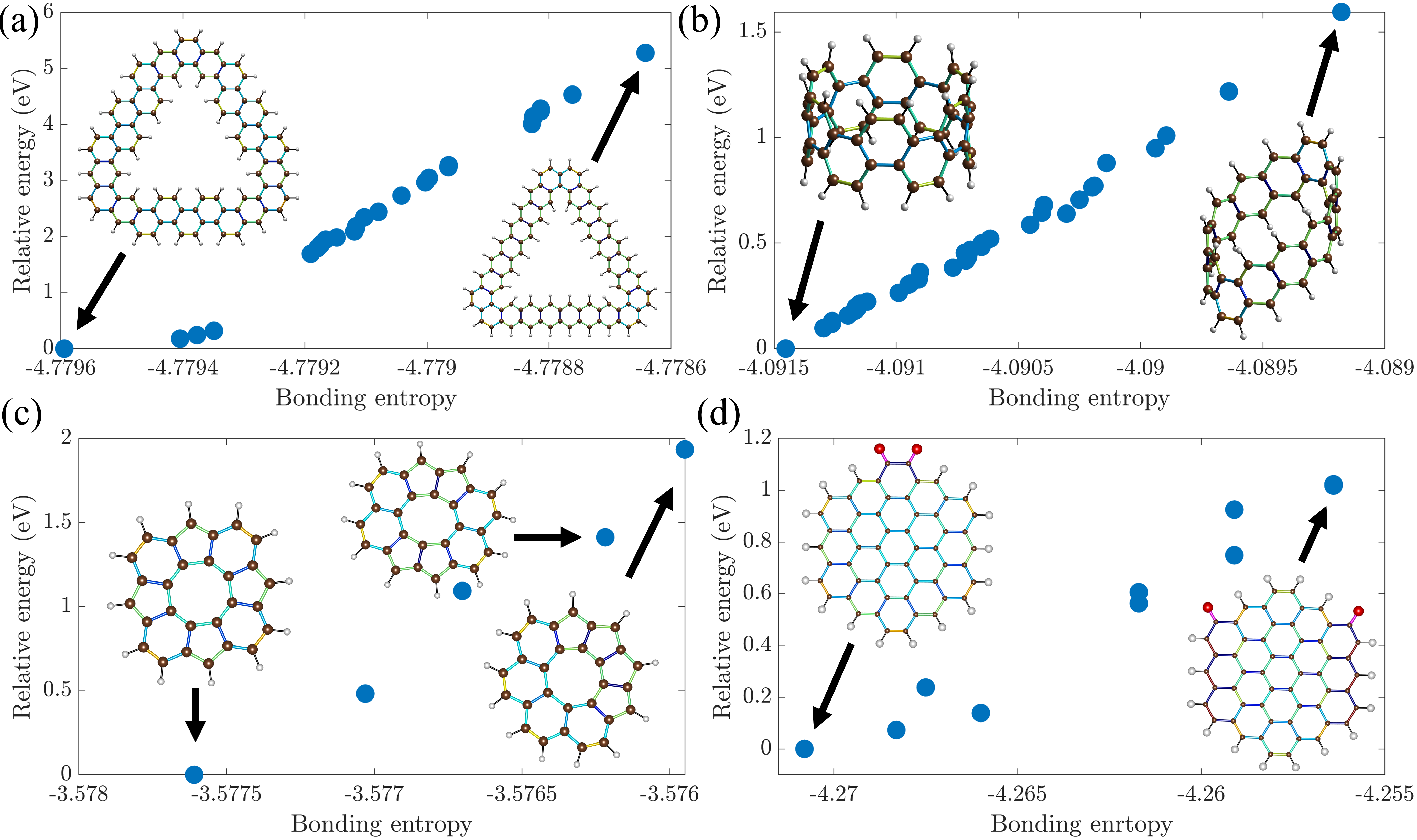}
    \caption{The relationship between relative  energy calculated by DFT  and bonding entropy for (a) cycloarenes; (b) carbon nanobelt; (c) carbon nanostructures with pentagon rings; (d)  C-H-O systems.  }
    \label{fig4}
\end{figure}

To validate the generality of the BFE model, we also considered other carbon-rich clusters  as shown in FIG. \ref{fig4}.   These molecules were evaluated using DFT calculations   \cite{PhysRevB.54.11169,PhysRevB.59.1758}, in which the stability of structures with low energy is confirmed by the high-level Heyd-Scuseria-Ernzerhof (HSE06)  calculation   \cite{10.1063/1.2187006,10.1063/1.1564060} as detailed in SM.   We take C$_{96}$H$_{48}$ with 24 hexagonal rings  as an example to demonstrate the predictive power in cycloarenes, which mainly consist of two types of structures: kekulene and clarene   \cite{C5CS00185D,C6CS00174B}. FIG. \ref{fig4}(a) clearly shows that the BFE has good linear consistence with the relative energy from DFT calculations. The most stable structure is marked in the left-top part of FIG. \ref{fig4}(a), in which the hexagonal rings stack with staggered alignment rather than arranged in a single row, since the ON of each bonds is more uniform to gain more bonding entropy. In contrast, the structure of highest energy in the right-bottom of FIG. \ref{fig4}(a) is composed of  three elongated PAH clusters,  losing more bonding entropy. FIG. \ref{fig4}(b) shows that the carbon nanobelt, which are rigid and thick segments of carbon nanotubes   \cite{doi:10.1126/science.aam8158},  can also be accurately predicted. Similar to cycloarenes, the most stable carbon nanobelt has the  hexagonal rings  in a staggered arrangement, while the most unstable nanobelt also has elongated PAH clusters. Therefore,  the elongated PAH clusters will remarkably contribute instability,  which originates from the  reactivate stability between anthracene and phenanthrene.

Next we focus on the carbon nanostructures with pentagon rings as shown in FIG. \ref{fig4}(c), in which the structure with isolated pentagons  has the lowest energy as predicted by BFE model and DFT calculations. The isolated pentagons can lead to a more evenly distribution of  ONs in the carbon nanostructure. The more adjacent pentagons there are, the less stable the structure becomes as depicted in FIG. \ref{fig4}(c). In addition to C-H systems, C-H-O systems can also be described by BFE model, where  the most stable structure occurs when the two oxygen atoms are in the ortho-position as shown in FIG. \ref{fig4}(d). Because the C-O is a double bond, then the bond distribution will be completely different with the introduction of oxygen atoms. Note that the carbon atom connected to oxygen atom will bonded two other carbon atoms by single bond, the ortho-oxygen atoms  will result in the entire system having only three C-C single bonds, thereby reducing one single bond, which  effectively increases the bonding entropy and  enhances the structural stability. Therefore, the BFE model can excellently predict the reactivate stability among widespread carbon nanostructures, which is compatible with previous theory.

With a given carbon nanostructure, there are numerous combinations that meet the octet rule, and  the charge density is attributed to all these combinations according to BFE model. The most likely electron distribution is deduced through ergodic theory, which entails averaging over all possible distributions   \cite{doi:10.1073/pnas.1421798112}.  The minimal  $F_{\textrm{b}}$ thus determines the optimal electron distribution with the maximum bonding entropy for a system, suggesting a uniform electron density to maximize resonance energy. Viewed as preliminary information, the number of the total electron constraint and octet rule aptly reflect the most informed current understanding, in line with the principle of  maximum entropy   \cite{ROWLINSON1970}.  Our BFE model emphasizes the important role of bonding entropy in the  carbon nanostructures, where the stabilities are dominated by delocalized electrons.

% It is crucial to note that while previous studies have combined DFT with model potentials to study structural stability in nanostructures,  massive {parameters} often require {training}. For example, methods such as the cluster expansion   \cite{doi:10.1021/ja9105623,doi:10.1021/acs.jcim.8b00413} and {universe} machine learning potentials   \cite{Fedik2022_nat_chem,doi:10.1021/acs.chemrev.0c01111} {may not} fully capture the actual interactions , writing complex interactions into massive model parameters, and obscuring the {chemical} origins of specific systems. In {this work}, we provide an analytic model to describe the interactions in PAHs, where the {behaviors} of electrons can be interpreted as a statistical ensemble average, particularly when bonding entropy is maximized. This allows the ground state electron configuration to represent the optimal resonance structure among all possible Lewis structures. 

\subsection{Electron structures prediction capability of BFE model}

\begin{figure*}
    \centering
    \includegraphics[width=\linewidth]{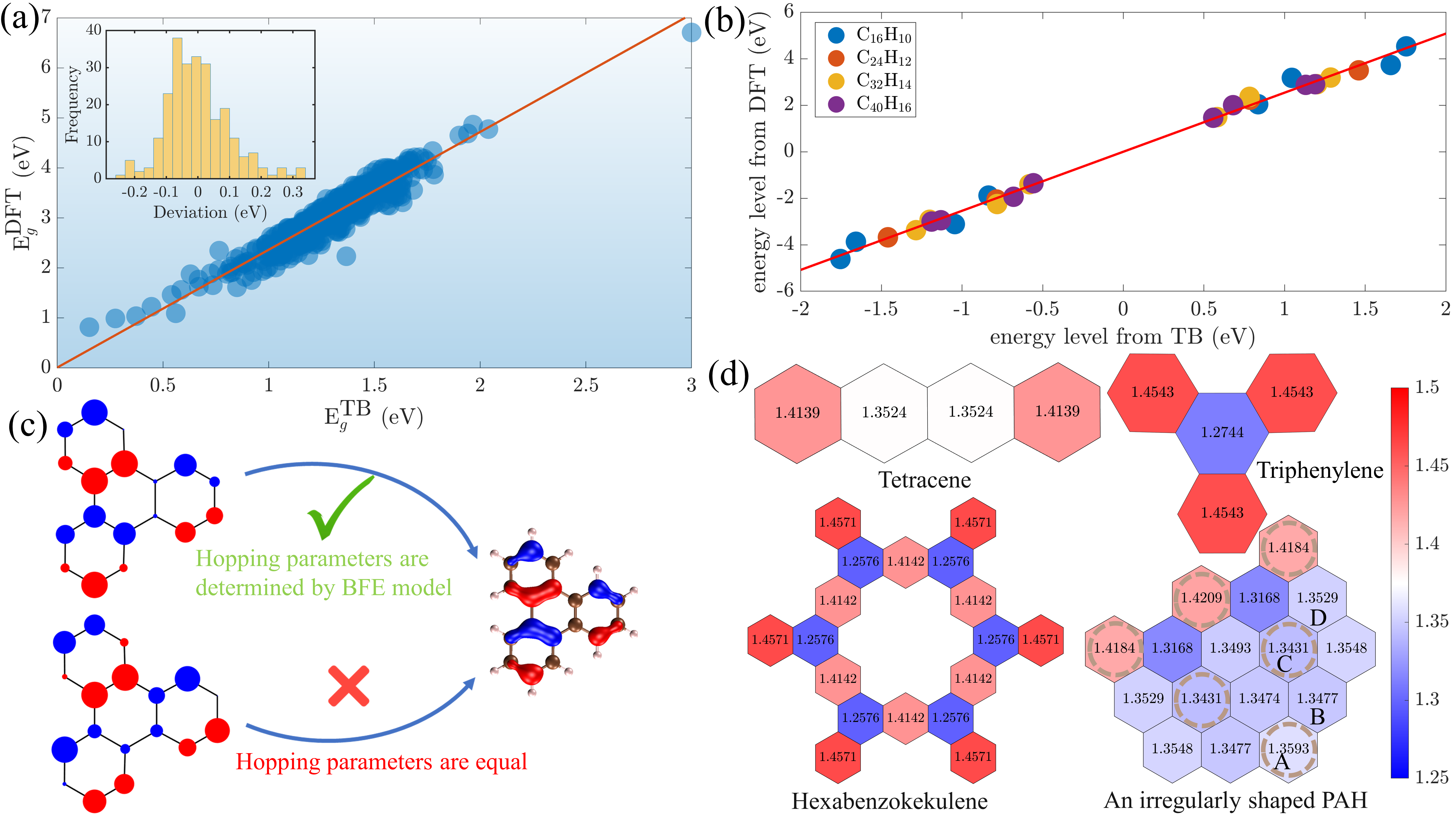}
    \caption{{(a) The linear relationship between the  energy gaps predicted by the TB method and those calculated by DFT, with the deviation distribution plotted in the inset.  (b) The  excellent agreement between  energy level predicted by TB model and DFT.  (c) The  molecular orbitals determined by the TB model (left panel) and DFT (right panel). (d) The defined local aromaticity for three representative molecules, where the color corresponds to the right colorbar.}}
    \label{fig5}
\end{figure*}

Due to the outstanding optoelectronic properties and applications in organic field-effect transistors   \cite{doi:10.1021/acs.chemrev.5b00312,doi:10.1021/jacs.2c04642,D3TC02876C}, the electronic structures of carbon nanostructures have attracted significant attention in both experimental and theoretical domains  \cite{doi:10.1021/jp021152e,doi:10.1021/acs.energyfuels.2c01170}. The TB model is typically employed  to describe the electronic structures, which often requires DFT calculations to fit integral parameters. Previous research indicates that accurately predicting energy gaps in PAHs requires up to thirteen hopping integral parameters   \cite{C7RA06891C}, essential for differentiating various C-C bonds. Herein, we assume the hopping parameter is proportional to the ON of the bond between carbon atoms $i$ and $j$, constructing the TB Hamiltonian for PAHs with nearest-neighbor couplings as follows:
\begin{equation}
    H =  -\sum_{i,j}t_{i,j}(c_i^{\dag}c_j+h.c.)
\end{equation}
where the on-site energy is set to zero, $t_{i,j}$ is the hopping integral, the subscript $i,j$ represents the indices of the nearest neighboring carbon atoms, and $c_i^{\dag}$, $c_j$ are the creation and annihilation operators of the $\pi$ electron at sites $i$ and $j$, respectively. The predicted energy gap by the TB model shows a linear correlation with the DFT-calculated energy gap in the HSE06 level, as depicted in FIG. \ref{fig3}(a). This relationship is expressed as $E_g^{\textrm{DFT}} = \gamma \times E_g^{\textrm{TB}} + E_{g}^0$, with parameters $\gamma=2.355$ {and} $E_g^0=0.01$ eV. Notably, over 70\% of the energy {gaps predicted by the TB model} deviate less than 0.1 eV from {those of the} DFT calculations as shown in the inset of FIG. \ref{fig5}(a), using only two external parameters, {namely $\gamma$ and $E_g^0$}. This underscores the efficacy of the ON in reflecting the hopping integral and bond strength. Additionally, the energy levels of the magic PAH clusters are also accurately predicted, as plotted in FIG. \ref{fig5}(b). Therefore, the ON of bonds serves as a reliable descriptor for predicting the electronic structures of PAHs, effectively capturing both structural stability and charge density.

To investigate  the wavefunction of  molecular orbitals,   we choose the highest occupied molecular orbital (HOMO)  of     triphenylene molecule to demonstrate the accuracy of  our TB model.   The  coefficients of each atomic orbitals  are the eigenvectors   corresponding to  the HOMO energy level as  shown in the upper left corner of  FIG. \ref{fig5}(c), where the radius of the circles represents the magnitude of the coefficients.  Note that red represents positive values while blue represents negative values, which is consistent with the  spatial distribution of wavefunctions calculated by DFT, as shown in the right panel of  FIG. \ref{fig5}(c). As a comparison,  the  bottom left part of  FIG. \ref{fig5}(c) shows the calculated molecular orbitals when all hopping integrals between the nearest neighbors are equal. Unfortunately,  the  magnitude of  coefficients on each C atom does not correspond to  the  calculations by DFT in the right panel of  FIG. \ref{fig5}(c), highlighting the importance of  distinguishing C-C bonds. It is  noteworthy that both the HOMO and the lowest unoccupied molecular orbital (LUMO) are  two-fold degenerate, where the four molecular orbitals are detailed in the SM.

As shown in FIG. \ref{fig5}(d), tetracene with a linear arrangement of acenes is less stable than triphenylene with Y-type acenes, and the energy gap of tetracene is smaller than that of triphenylene   \cite{ https://doi.org/10.1002/chem.201905087}. Here we define the mean value $\kappa$ of ONs on six-membered rings to describe the local aromaticity. We can assert that the ring is akin to the ``fully-aromatic" benzene if $\kappa$ approaches 1.5; otherwise,  it exhibits less aromaticity. It is apparent that $\kappa$ values in triphenylene are larger than those in tetracene, which suggests additional aromaticity and stability   \cite{doi:10.1021/cr9903656}. The defined $\kappa$ offers an avenue to describe the aromaticity, providing insight into the electronic properties of PAHs. Furthermore, triphenylene can self-assemble into a hexabenzokekulene with six-fold symmetry   \cite{doi:10.1021/ci00067a002}, which has the largest energy gap among all isomers. Note that the hexabenzokekulene holds 12 highly aromatic rings with red coloring of the 18 hexagonal rings, in which a particularly high proportion of aromatic rings results in a larger band gap and high stability.   FIG. 5(d) shows the large irregularly shaped PAHs as the  Clar’s  rule-breaking examples     \cite{doi:10.1021/acs.jpca.6b00972},  where the Clar's rule shows that the A, C rings are  more aromatic  than ring B, D. However,  according to several aromaticity descriptors on the rings (MCI$^{1/n}$, HOMA, and NICS(1)$_{zz}$), the aromaticity of   ring A-D  is comparable   \cite{Wang2022_jpca}, which is also in good agreement with  the local aromaticity description in our parameter-free model.

For $sp^2$ bonded carbon  nanoclusters with 2c-2e bonds, the combined use of the BFE and the octet rule has been effective in predicting various properties.  We have extended this approach to $sp$-$sp^2$ hybridized graphyne structures  \cite{PhysRevLett.108.086804,PhysRevLett.108.086804,PhysRevB.86.115435}, showing the capability of BFE in forecasting structural properties in periodic systems. Unlike isolated PAH structures, periodic boundary conditions needs to be considered in the unit cell for graphyne. We have systematically generated a range of nonequivalent candidates for $sp$-$sp^2$ hybridized carbon allotropes, covering numbers of carbon atoms from 12 to 60   \cite{PhysRevMaterials.7.084202}. The definition of BFE for graphyne follows Eq. \ref{free_BE}, where the globally minimal BFE corresponds to the bonding energy.  Consequently, the BFE can elucidate structural stability in both molecular and periodic systems, highlighting its wide-ranging applicability.

\begin{figure}[ht]
    \centering
    \includegraphics[width=\linewidth]{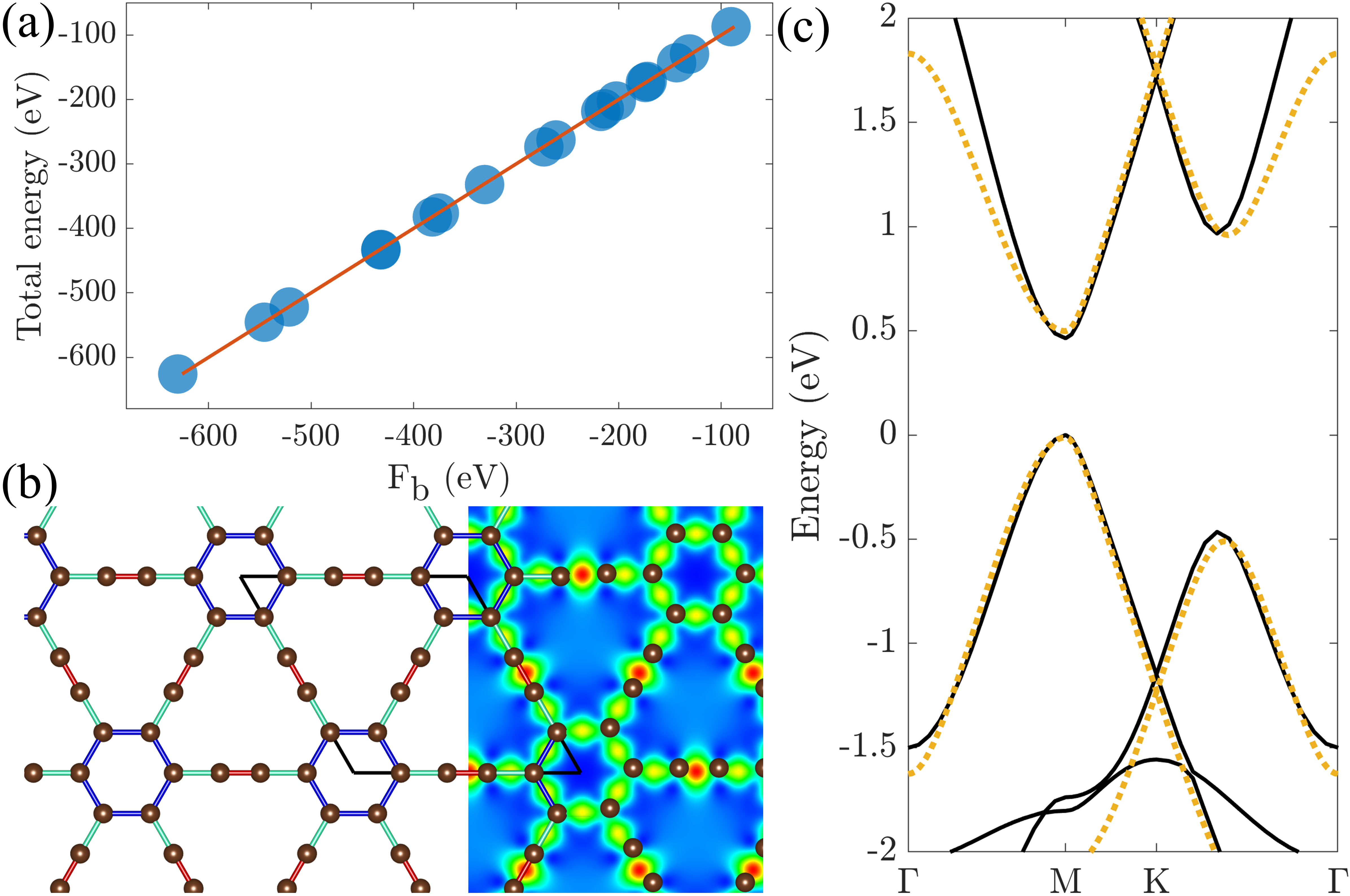}
    \caption{(a) The linear correlation between BFE and bonding energy derived from DFT. (b) Distribution of  ON  and charge density  for the $\gamma$-graphyne structure. (c) The band structure determined by the DFT method (solid line) and the TB method (dashed line).}
    \label{fig6}
\end{figure}

Moving beyond structural stability, we focus on the electronic structures of graphynes. For the synthesized $\gamma$-graphyne   \cite{Hu2022}, we present the predicted ON of each bond alongside the charge density in Fig. \ref{fig6}(b). It is observed that bonds between two-coordinate carbon atoms are stronger than those between three-coordinate carbon atoms, as indicated by the higher ON and the red area in the charge density distribution. By optimizing ON of each bond to minimize BFE, we construct a TB model for graphyne, contributed by $p_z$ electrons, to analyze electronic structures. In this model, the hopping integral is proportional to the exponential of the ON. The TB-based band structures (dashed line) as given in Fig \ref{fig6}(c), which show considerable agreement with DFT results (solid line) near the Fermi level. A single proportionality coefficient is fine-tuned to match the band gap determined by DFT. This matching process underscores the effectiveness of ON in differentiating  hopping integrals between nearest carbon atoms.  The band structure predictions for  other graphyne structures are further detailed in FIG. S4 of  SM. Here, the optimal  ON for each bond serves as a reliable measure of the bond strength and effectively describes the electronic structures of various dimensional systems.

\section{Conclusions}
In conclusion, for the first time, we have developed a parameter-free statistical model that determines the electron density and related properties in carbon nanostructures across various dimensions. Without relying on pre-training data from DFT, the bonding free energy defined in our model shows a linear correlation with the bonding energy derived from DFT. This highlights the crucial role of bonding entropy in stabilizing carbon nanostructures with delocalized electrons. The optimal occupancy number of each bond accurately reflects bond strength and closely correlates with bond length and charge density distribution, which also serves as the basis for determining hopping integrals when constructing TB models. These TB models have proven effective in predicting electronic structures for various dimensional carbon nanostructures, thus enriching our understanding of electronic behaviors. Furthermore, the mean value of the occupancy number offers a measure of local aromaticity for six-membered rings, aligning well with current calculations for aromaticity. By incorporating bonding free energy, we provide a statistical perspective to understand the quantum behaviors of electrons, offering a concise yet precise approach for determining structural stability and electronic properties in carbon nanoclusters and graphynes, which may also hold potential for broad applicability in other systems as well. 

\section{SUPPLEMENTARY MATERIAL}
The supplementary material contains information of the computational methods,  the detaild derivation of bonding free energy model, the energy level and aromaticity prediciton of PAH molecules and the band structure prediciton of graphynes.

\section{AUTHOR DECLARATIONS}
\subsection*{Conflict of Interest}
The authors have no conflicts to disclose.

\begin{acknowledgments}
This work is supported by Guangdong Basic and Applied Basic Research Foundation (Grant No.  2023A1515110894 and No.2023A1515010734), Guangdong Provincial Key Laboratory of Functional and Intelligent Hybrid Materials and Devices (Grant No. 2023-GDKLFIHMD-04),  the National Natural Science Foundation of China (Grant No. 12374182 and No. 12204224) and the Shenzhen Science and Technology Program (Grant No. RCYX20200714114523069).
\end{acknowledgments}

% \bibliography{apssamp}


\end{document}